\documentclass[12pt]{article}
\usepackage{graphicx,amssymb}
\usepackage[figuresright]{rotating}

\usepackage{epsfig}
\usepackage{graphicx}

\newcommand{\tr}{\mathop{\rm Tr}}

\newcommand{\beq}{\begin{equation}}
\newcommand{\eeq}{\end{equation}}
\newcommand{\bea}{\begin{eqnarray}}
\newcommand{\ena}{\end{eqnarray}}

\begin{document}
\sloppy \vspace{-1cm}

\begin{flushright}
{ITEP-LAT/2006-07}, \vspace{0.2cm}
{HU-EP-06/17}
\end{flushright}
\begin{center}
{\baselineskip=24pt {\Large \bf
Once more on the interrelation between Abelian monopoles and P-vortices in
$SU(2)$ LGT
%---------------------------------------------------------------------
}\\
\vspace{1cm}

{\large P.~Yu.~Boyko$^{\dag}$,
        V.~G.~Bornyakov$^{\dag,*}$,
    E.-M.~Ilgenfritz$^{**}$,
        A.~V.~Kovalenko$^{\dag}$,
    B.~V.~Martemyanov$^{\dag}$,
    M.~M\"uller-Preussker$^{**}$,
        M.~I.~Polikarpov$^{\dag}$ and
    A.~I.~Veselov$^{\dag}$ }

\vspace{.5cm} {\baselineskip=16pt { \it
$^{\dag}$ Institute of Theoretical and  Experimental Physics,\\
B.~Cheremushkinskaya~25, Moscow, 117259, Russia\\
$^{*}$ Institute for High Energy Physics, \\
Protvino, 142281, Russia\\
$^{**}$ Institut f\"ur Physik, Humboldt-Universit\"at, \\
Newtonstrasse 15, 12489 Berlin, Germany} }}

\end{center}
\abstract{
%---------
We study the properties of configurations from which P-vortices
on one hand or Abelian monopoles on the other hand have been removed.
We confirm the loss of confinement in both cases and investigate in what 
respect the modified ensembles differ from the confining ones from the 
point of view of the complementary confinement scenario.
}
\date{July 5, 2006}
\newpage

%-----------------------------------------------

\section{Introduction}
%----------------------
\label{sec:Introduction}

There are two popular phenomenological explanations of confinement in lattice
gluodynamics: the monopole~\cite{monopole} and the center 
vortex~\cite{greensite_1,greensite_2,greensite_3,langfeld_1,langfeld_2,langfeld_3,cornwall}
confinement mechanism that have been critically discussed in Ref.~\cite{reviews}.
Monopoles and center vortices are defined by projection to $U(1)^{N-1}$ or $Z(N)$
gauge fields, respectively. Hence, the center vortices are called P-vortices in 
order to emphasize the difference from extended (continuum) 
vortices~\footnote{We will interchangeably use both terms in the following.}.
These two types of excitations reproduce, respectively, about 90\% \cite{simann} 
and about 70\% \cite{drama} of the non-Abelian string tension and are able to 
explain other nonperturbative properties. Therefore, removing monopole 
fields~\cite{miyamura_1,miyamura_2} 
or P-vortices~\cite{deforcrand} should leave only inert 
gauge field configurations, unable to confine external charges or to break chiral 
symmetry.

It was very attractive to conjecture~\cite{conjecture,greensite_4} that
monopoles and P-vortices are geometrically interrelated. Indeed, this was
found to be the case in $SU(2)$ gluodynamics: more than 90\% of monopole
currents are localized on the P-vortices~\cite{greensite_4,interplay,kovpolsyrzakh}.
One of our goals in the present paper is to elaborate on this interrelation
between monopoles and vortices.

In Ref.~\cite{deforcrand} the operation of center vortex (P-vortex) removal has 
been invented, and it was demonstrated that the quark condensate and the topological 
charge are destroyed thereby. Later it has been found~\cite{z2landau} that the 
number of links that have to be modified in this operation can be essentially 
reduced when the $Z(2)$ gauge freedom is used to minimize the number of negative 
trace links. The remaining negative links can be interpreted in terms of 
three-dimensional domains.

The operation of monopole removal has been invented by Miyamura et al. in
Ref.~\cite{miyamura_1,miyamura_2}. It has been shown that not only the quark 
condensate and the topology is destroyed thereby, but that the Abelian monopole 
field alone, completed again by the non-diagonal gluons, carries all necessary 
degrees of freedom in order to allow the reconstruction of chiral symmetry 
breaking and topological charge. 

All technical details (including those of gauge fixing) relevant for our study 
are relegated to the Appendix.

Ph. de Forcrand and M. D'Elia~\cite{deforcrand} were the first to ask what 
happens to the Abelian
monopoles when P-vortices are removed. We have started our work from the opposite 
question: what remains from the center vortex degrees of freedom when Abelian 
monopoles are subtracted from the Abelian part of the gauge fields. Originally, 
it was only for completeness that we also repeated the study of 
Ref.~\cite{deforcrand}.
Not unexpectedly, we found that the results depend on the method of gauge fixing
and details of its algorithmic implementation. But turning again to de Forcrand's
and D'Elia's
problem with the perfectionated gauge fixing has resulted in a paradox that has
led us to a closer investigation what properties the network of monopole world
lines must possess in order to confine.

Most of the above mentioned results in the literature was obtained for the
Direct Center Gauge (DCG)~\cite{greensite_3} and for the Indirect Center Gauge
(ICG)~\cite{greensite_1}. In the present paper we also use both of them. 
The Laplacian center gauge \cite{deForcrand0008016}, also used for studies of 
the P-vortices, is not studied here. Although the absence of the Gribov 
ambiguity is a virtue of this gauge, we consider the absence of scaling
of the resulting density of P-vortices~\cite{langfeld0101010} as a less 
appealing feature. We have been doing all gauge fixings by means of simulated 
annealing, repeating however the procedure for a number of copies. In this 
respect we have done our best (according to the present state of the art) 
to circumvent the infamous Gribov ambiguity.

In all studies described in this paper we have used a confining ensemble 
of 100 configurations on a $24^3\times 6$ lattice, generated at $\beta=2.35$ 
using the Wilson action. 
This corresponds to a lattice spacing given by 
$a~\sqrt{\sigma} = 0.3108(17)$~\cite{LT2001}\footnote{We thank Mike Teper for 
confirming that this value actually refers to $\beta=2.35$.}, in other words, 
to a temperature 
$T/T_c = 0.7562(80)$, where the ratio $T_c/\sqrt{\sigma} = 0.7091(36)$ 
(obtained for $SU(2)$ gluodynamics with Wilson action in Ref.~\cite{LTW2003}) 
has been used. Adopting the standard value of the string tension, 
$\sigma = (440 \mathrm{MeV})^2$, the estimated lattice spacing would be 
$a=0.139(1)$ fm, resulting in a spatial volume of $(3.336~\mathrm{fm})^3$.

In order to demonstrate the confining property or the loss of confinement
in the case of finite temperature adopted here, we use the correlator of 
the Polyakov loop,
\begin{equation}
a V(r) = - \frac{1}{N_{\tau}}~\log \langle P({\vec x}) P^{*}({\vec y}) \rangle
\end{equation}
with $r = |{\vec x} - {\vec y}|$. Strictly speaking, this is the excess free 
energy related to the presence of a quark-antiquark pair. For simplicity we 
will call it potential, however. For the full non-Abelian potential the 
Polyakov loop
\begin{equation}
P({\vec x}) = \frac{1}{2} \tr \prod_{t=1}^{N_{\tau}} U_{{\vec x},t;4} 
\end{equation}
is used. When the monopole-generated potential shall be sorted out, the 
monopole part of the Abelian Polyakov loop, 
\begin{equation}
P^{mon}({\vec x})= \mathrm{Re} \exp
\left( \sum_{t=1}^{N_{\tau}} i \theta^{mon}_{{\vec x},t,4} \right) 
= \cos \left( \sum_{t=1}^{N_{\tau}} \theta^{mon}_{{\vec x},t,4} \right) \; , 
\label{eq:monopole-P}
\end{equation}
is entering the correlator. The monopole part of the gauge field 
$\theta^{mon}_{x,\mu}$ is defined in eq. (\ref{eq:monfield}) in the Appendix.
Finally, in order to measure the vortex contribution to the potential, 
the Polyakov loop is constructed in terms of the $Z(2)$ projected links
\begin{equation}
P^{vort}({\vec x})=  \prod_{t=1}^{N_{\tau}} Z_{{\vec x},t;4} \; ,
\label{eq:vortex-P}
\end{equation}
with $Z_{x,\mu}$ defined in eq. (\ref{eq:Zdef}) in the Appendix.

As already said, in the course of the present paper, we are studying monopole
currents in configurations where the P-vortices are removed and P-vortices in 
configurations where the monopoles are removed. The results are trivial for the 
case of ICG, as we will show in Section~\ref{sec:ICG}. Here, only nonpercolating 
monopole currents or only small, nonpercolating center vortices, respectively, 
are left over.

Note that in the Laplacian center gauge the monopoles are 
located on the center vortices \cite{deForcrand0008016} and thus removal 
of the vortices automatically leads to removal of the monopoles. 
We expect that, similar to ICG, removal of monopoles in this gauge will 
give rise to nonpercolating center vortices.

In the case of DCG, when the emerging center vortices are removed, the situation 
is more challenging (as described in Section~\ref{sec:DCG_nocenter}). We observed 
that removing P-vortices leads to gauge field configurations with monopole currents 
even more dense and still percolating although confinement is lost. This is possible
since monopole and vortex degrees of freedom are, at first sight, only loosely 
connected in DCG, in contrast to ICG. This intriguing observation has prompted us 
to search for the necessary conditions for monopole percolation to generate 
confinement. We found the reason why monopole percolation, although necessary, 
is not sufficient to create a non-vanishing string tension. We shall argue below 
that this effect is due to a very special structure of the percolating monopole 
cluster: it is highly correlated at small distances which screens monopole currents 
more than this is the case in normal configurations.

Finally, in Section~\ref{sec:DCG_nomono}, we analyse how the subtraction of
singular components in the Abelian gauge field (generated by monopoles) changes
the P-vortex structure. In Section~\ref{sec:Conclusions} we draw some conclusions.

\section{Monopole and vortex percolation}
%------------------------------
\label{sec:percolation}

The paradox mentioned in the Introduction is the observation that monopole 
percolation is not sufficient to guarantee confinement. This is surprising
because monopole percolation has become synonymous for monopole condensation.
The monopole condensation mechanism of confinement has been, together with
the Dual Superconductor picture, proposed by 't Hooft~\cite{tHooft1} and 
Mandelstam~\cite{Mandelstam} implying that long-range physics is dominated
by Abelian degrees of freedom (Abelian dominance). Gauge fixing and Abelian 
projection have been suggested~\cite{tHooft2} as a method to verify this idea. 
On the lattice this has been first implemented in Ref.~\cite{Wiese}. 

When Abelian dominance of the string tension had been verified on the 
lattice~\cite{AbelDomin}, this has boosted the development in two directions.
At first, several observables have been constructed in terms of the monopole 
currents (which are immediately available in the projected theory) serving
the intention that each of them should reflect the phenomenon of monopole 
condensation. In this paper we will not comment on the second direction, the
construction of a true monopole condensation disorder parameter, following the 
early proposals in Refs.~\cite{Frohlich1,DiGiacomo1}. This observable should have  
a non-vanishing expectation value in the confining phase (hopefully independently 
of the particular Abelian projection~\cite{tHooft2}) indicating the spontaneous
breakdown of the corresponding magnetic $U(1)$ 
symmetry of the latter~\cite{disorderparameter,frohlich2,monop_condense}.

Returning to the direct monopole observables in MAG, the first proposal was to 
measure the monopole current density $\rho_{\rm mon}$ and to establish its scaling 
property with respect to $\beta$ at zero temperature~\cite{density}. 
The monopole density and similarly the vortex density are defined as
\begin{equation}
\rho_{\rm mon}=\frac{<N_{\rm mon}>}{4 N_s^{3} N_{\tau}} \quad \mathrm{and} \quad
\rho_{\rm vort}=\frac{<N_{\rm vort}>}{6 N_s^{3} N_{\tau}} \; ,
\label{eq:density_def}
\end{equation}
where $N_{\rm mon}$ is the number of dual links occupied by monopole currents
and $N_{\rm vort}$ the number of dual plaquettes belonging to the vortex area.

The observed scaling of the monopole density has later found to be premature. 
Moreover, the density lacks scaling and universality~\cite{BornMMP,Universality}
and requires to be separated into ``ultraviolet'' and ``infrared component'',
the scaling behavior of which is different, meaning that the UV component reflects
only short distance artefacts without relevance for confinement. Only the IR
component turned out to be scaling and relevant for condensation~\cite{ivanenko} 
and for confinement. Relatively soon after the first quantitative study of 
$\rho_{\rm mon}$~\cite{density}, the focus has turned to structural properties 
of the network of monopole currents, which has found to be clearly undergoing a 
change at the deconfinement phase transition~\cite{BornMitrMMP}. The change 
concerned the distribution of connected clusters with respect to the length $L$ 
({\it i.e.} the number of participating monopole currents). 
This distribution is denoted as $N(L)$ and counts the number of connected clusters
of given size $L$ in the ensemble of configurations. In \cite{BornMitrMMP}, 
and later in Refs.  \cite{Suzuki-clust} and \cite{HartTeper_1,HartTeper_2} it has 
been clearly demonstrated that, in the confinement phase, each configuration 
contains a single macroscopic cluster (in modern parlance the ``infrared monopole 
component'') of many connected dual links carrying non-vanishing monopole current, 
apart from a big number of small clusters (the ``ultraviolet monopole component''). 
The macroscopic clusters were found to be {\it percolating}. This is meant in the 
sense that they are impossible to enclose in a 4D cuboid smaller than the periodic 
lattice in one ore more directions or in the sense of a cluster two-point 
function~\cite{ivanenko} not to vanish at ``infinite'' distance. Obviously, this 
notion of percolation does not take into recognition the orientation of the 
monopole currents. The eventual occurrence of more than one macroscopic cluster 
was found to be an effect of smaller volume.  

Of course, it is not difficult just to evaluate Wilson loops under the influence 
of the Abelian field due to all monopole currents~\cite{monopole,Stack_1,Stack_2}, 
a procedure that clearly takes the orientation into account. In this way, monopole 
dominance~\cite{monopole,simann} has been demonstrated. Later it has been found that
the macroscopic clusters (and only these) build up the Abelian string 
tension~\cite{HartTeper_2}. The essence of the deconfinement transition in this 
light had already earlier been identified~\cite{BornMitrMMP} as the decomposition 
of the macroscopic cluster into many smaller ones, {\it i.e.} a qualitative change 
of the size distribution of the clusters, accompanied by the generation of an 
anisotropy of the current density. In the Euclidean time direction, the percolation 
property of the remaining larger clusters persits above $T_c$, whereas there is
no percolation anymore in the spatial directions. 

All this contributed to the common belief that spacelike percolation is a 
necessary {\it and} sufficient condition for confinement. In this paper we find 
an interesting counter example showing that a percolating cluster does not
necessarily confine, as the direct evaluation of the monopole contribution to 
the Wilson loops reveals.

In the case of the P-vortex mechanism, the direct evaluation of their contribution
to Wilson loops (by a factor $(-1)^{\rm linking~number}$ for each vortex), the
fact of center dominance (of the string tension) and the scaling property of the 
P-vortex density $\rho_{\rm vort}$ (being equal to the density of negative 
plaquettes) have been emphasized from the very beginning~\cite{langfeld_1}. 
Also here, the existence of one macroscopic P-vortex, 
a closed surface consisting of many dual plaquettes, was found to be a prerequisite 
of confinement, and this P-vortex itself was found percolating~\cite{greensite_5} 
in all directions. The deconfinement transition was later identified, in analogy
to the monopole mechanism, with a de-percolation 
transition~\cite{realsuperconductor,langfeld_3,Engelhardt}, 
{\it i.e.} with the macroscopic vortex 
being replaced by smaller ones. If big enough, these are percolating in the time 
direction, but no more in spatial directions. 

In our investigation, for P-vortices we did not find any indication that vortices 
would persist to be spatially percolating in the case confinement has been destroyed 
({\it e.g.} by the removal of monopoles). 

\section{Indirect center gauge}
%------------------------------
\label{sec:ICG}

ICG fixing includes two steps. First, the maximally Abelian gauge (MAG) is
fixed and Abelian projection is made. Then the Abelian gauge fields (and the
remaining Abelian gauge invariance) are used to fix the (Abelian) maximal 
center gauge. After the center projection is made, the emerging $Z(2)$ 
configuration is looked up for negative links $Z_{x,\mu}=-1$ (see Appendix 
for further details). In Fig.~\ref{fig:cl_loop_vri} the length spectrum 
$N(L)$ of the monopole clusters, obtained after fixing the MAG, is compared 
with the same spectrum after the P-vortices have been removed. 
More precisely, here and later $N(L)$ represents the number of clusters 
of size $L$ found in all 100 configurations. Note, that the vortex 
removal operation consists in changing the sign of links with negative 
$\cos\left( \theta_{x,\mu} \right)$ in the (Abelian) maximal center gauge, 
and that this operation preserves the MAG \cite{simann}.

The original monopole density (per dual link) is 
$\rho_{\rm mon} = 0.0357 \pm 0.0002$ and the original vortex (area) density 
(per dual plaquette) is $\rho_{\rm vort} = 0.0805 \pm 0.0002$.

It is seen that the percolating monopole clusters with length around 7,000
that are present before in each lattice configuration, are absent from the 
length spectrum after the vortices have been removed. Only small monopole
clusters survive. The density of monopoles $\rho_{\rm mon}$ is reduced by a
factor $0.500 \pm 0.003$.  
Combining this observation with the above-mentioned fact of strong correlation 
between monopoles in MAG and vortices in ICG we conclude that the removal of 
P-vortices gives rise to removal of almost all monopoles belonging to 
the infrared part of the monopoles. It is well known that the string tension 
vanishes in the modified ensemble.
%-------------------------------------------------------------------------------
\begin{figure}
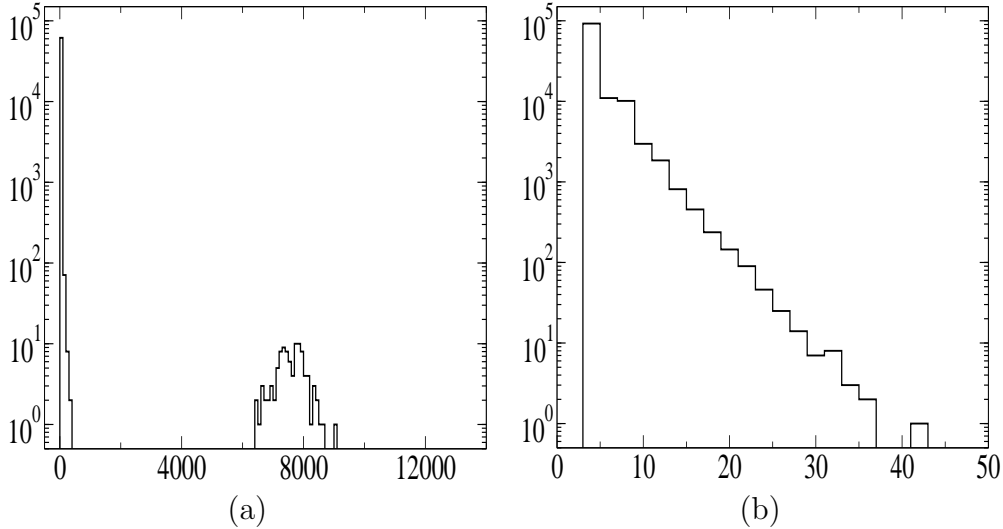

\begin{center}
\begin{tabular}{cc}
\includegraphics[width=.4\textwidth,height=.4\textwidth]{monclequi_1.eps}&
\includegraphics[width=.4\textwidth,height=.4\textwidth]{idcgcl.eps}\\
(a) & (b)
\end{tabular}
\caption{The monopole clusters length distribution $N(L)$ for Wilson $\beta=2.35$
on $24^3 6$ lattices before (a) and after (b) removing the center vortices found 
according to ICG.}
\label{fig:cl_loop_vri}
\end{center}
\end{figure}
 %---------------------------------------------------------------------

\begin{figure}
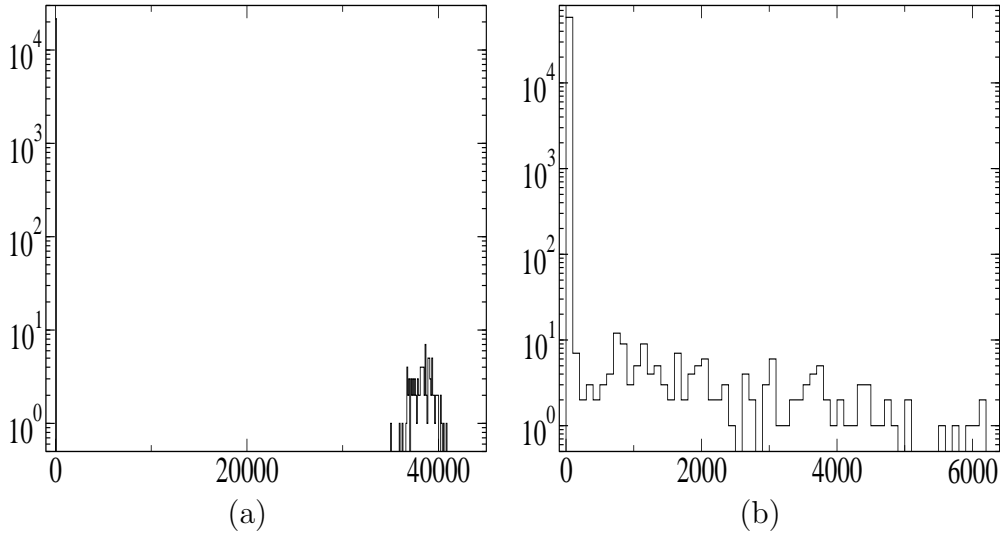

\begin{center}
\begin{tabular}{cc}
\includegraphics[width=.4\textwidth,height=.4\textwidth]{idcgvorclequi.eps}&
\includegraphics[width=.4\textwidth,height=.4\textwidth]{idcgvorclmonrem.eps}\\
(a) & (b)
\end{tabular}
\caption{The area spectrum of center vortices for Wilson $\beta=2.35$
on $24^3\times 6$ lattices (a) for the original ensemble put into ICG,
and (b) after monopoles have been removed.}
\label{fig:vort_sp}
\end{center}
\end{figure}

In Fig.~\ref{fig:vort_sp} the area spectrum of P-vortices, as it is found
after ICG of the original ensemble, is compared with the spectrum that is
obtained when the singular (monopole) part is subtracted from the Abelian
projected gauge field after MAG, before the ICG is finally accomplished. 
Thereby the original vortex density $\rho_{\rm vort}= 0.0805 \pm 0.0002$ 
is reduced to one fifth, $\rho_{\rm vort}= 0.0162 \pm 0.0003$, without 
monopoles. 

As long as the monopole fields are retained there is one big percolating vortex
with an area of about $40,000$ plaquettes present in each configuration 
(in addition to a lot of small size ones, with area $\lesssim 100$). 
After the monopole 
contributions to the Abelian field are removed the percolating P-vortex has 
disappeared from all configurations, the largest vortices being smaller in 
area than $6,000$ plaquettes. Thus we observe that removing monopoles 
does not lead to the loss of all big P-vortices (which would be analogous to 
what happens to the monopoles after vortex removal), but rather to splitting 
of the single percolating P-vortex into a number of still relatively extended
P-vortices. 

%---------------------------------------------------------------------
%---------------------------------------------------------------------

The above results are in agreement with both scenarios of confinement, claiming
that a non-vanishing string tension is due to percolating clusters of
monopoles~\cite{HartTeper_1,HartTeper_2} 
or due to the existence of percolating center 
vortices~\cite{greensite_5}, respectively.

\section{Direct center gauge: removing center vortices}
%------------------------------------------------------
\label{sec:DCG_nocenter}

When DCG fixing and center projection is completed, the vortex density
is $\rho_{\rm vort} = 0.0605 \pm 0.0002$.
As we mentioned already in the Introduction we found something unexpected
for the case of DCG, when center vortices are removed from the configurations.
This is done in the following way. Each Monte Carlo configuration is gauge-fixed
twice, to the MAG and to the maximal center gauge by DCG. The non-Abelian gauge
field is then projected either to an Abelian $U(1)$ gauge field (from MAG) or 
to an $Z(2)$ gauge field (from DCG). 
In DCG, the remaining $Z(2)$ gauge freedom is 
exploited to reduce the number of negative ($Z_{x,\mu}=-1$) links.

In Ref.\cite{deforcrand}, de Forcrand and D'Elia noticed already the existence 
of a few
percolating monopole clusters after the removal of vortices, clearly reduced
in comparison to the amount of huge monopole clusters in the normal case.
He mentioned, that only a handful of large clusters survives, whereas most of
them are broken into pieces. Moreover, even the remaining largest clusters,
according to his observation, did not contain monopole loops that wind around 
the periodic lattice. For the case of asymmetric lattices (finite temperature) 
considered in our work this observation would imply the absence of spatial 
winding. Qualitatively our results are in agreement with this 
expectation. Here we do not consider the winding number of the emerging monopole 
network clusters. Instead, we shall apply the decomposition of the clusters into 
non-intersecting monopole loops and find that none of them is spatially winding 
around the lattice.

Note that certain quantitative differences are possible due to the fact that 
DCG and MAG both are afflicted by their Gribov ambiguities. Our method differs 
from the method in Ref.~\cite{deforcrand} that we {\em (i)} always used the 
simulated annealing version of the gauge fixing procedure in question, 
{\it i.e.} we obtained higher local maxima for respective gauge fixing functionals,
and {\em (ii)} the $Z(2)$ Landau gauge is realized. Both modifications are expected 
to lead to a minimal number of non-vanishing monopole currents and a minimal number 
of negative $Z(2)$ links, respectively. The realization of the $Z(2)$ Landau gauge 
has no effect on the monopole configuration.
The procedure of P-vortex removal then requires the multiplication of each 
link by $Z_{x,\mu}$. This is applied not only to the original non-Abelian
configuration but also to the Abelian projected configuration. In this 
way the monopole content of configurations with P-vortices removed can be 
easily compared with the monopole content of the original configuration.
The monopole density $\rho_{\rm mon}$ is enhanced to 
$\rho_{\rm mon}=0.0551 \pm 0.0003$.
We found, in particular, that the monopole clusters still contain a percolating 
component in agreement with Ref.~\cite{deforcrand}.

\begin{figure}
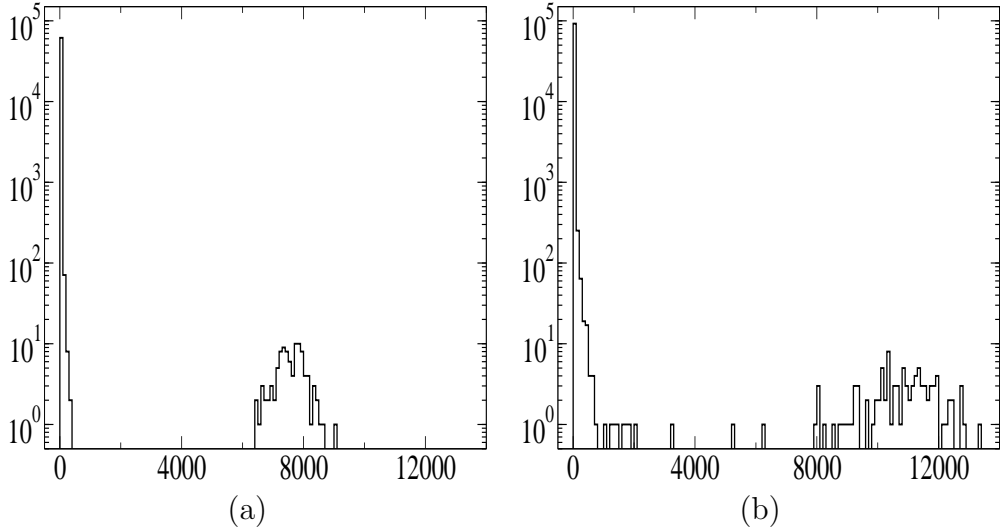

%% [hpbt]
\begin{center}
\begin{tabular}{cc}
\includegraphics[width=.4\textwidth,height=.4\textwidth]{monclequi_2.eps}&
\includegraphics[width=.4\textwidth,height=.4\textwidth]{dcgsacl.eps}\\
(a) & (b)
\end{tabular}
\caption{The same as Fig. \ref{fig:cl_loop_vri}, but (b) shows the situation 
after the center vortices found according to DCG have been removed.}
\label{fig:DCGmonspectr}
\end{center}
\end{figure}

\begin{figure}
\begin{center}
\begin{tabular}{cc}
\includegraphics[width=.45\textwidth,angle=0]{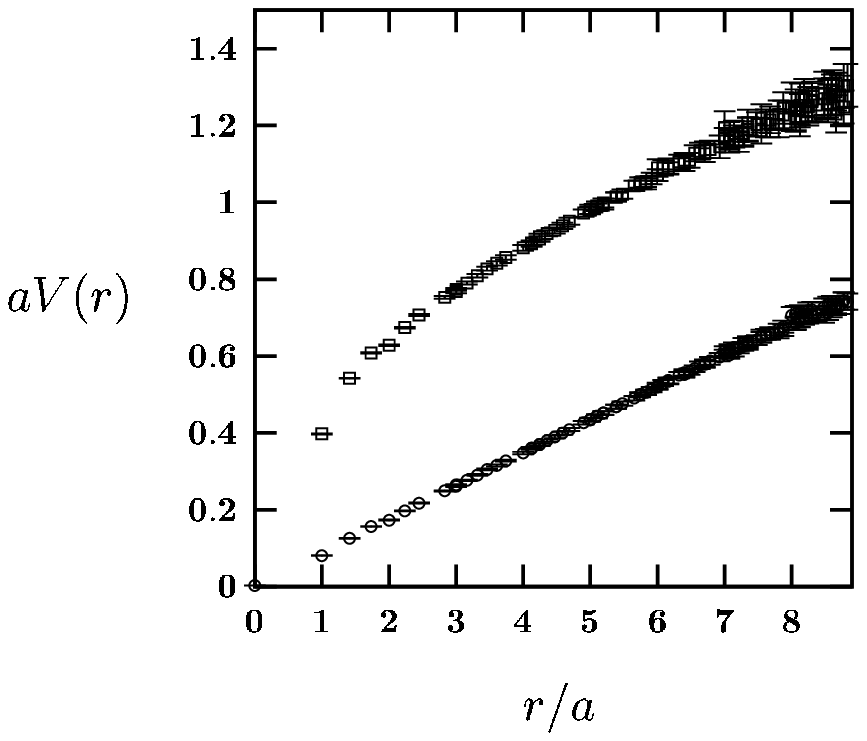}&
\includegraphics[width=.48\textwidth,angle=0]{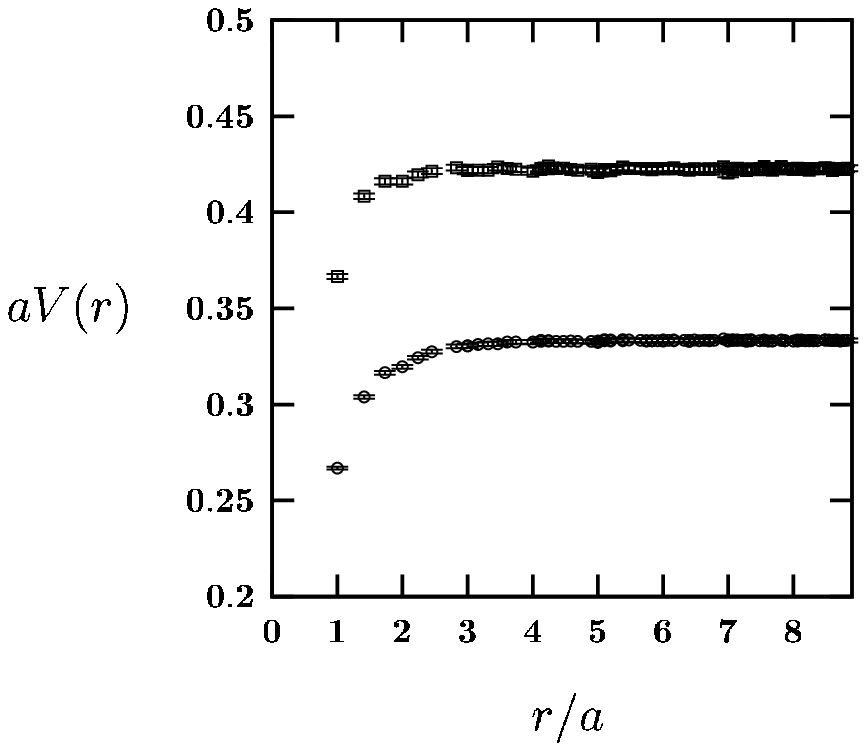}\\
(a) & (b)
\end{tabular}
\caption{
The non-Abelian potential ($aV(r)$, denoted by squares) and
the monopole potential ($aV_{mon}(r)$, denoted by circles)
at  $\beta=2.35$ on $24^3\times 6$ lattices
before (a) and after (b) removing of center vortices.}
\label{fig:fullmonabpotentials}
\end{center}
\end{figure}

The apparent paradox is demonstrated in Fig.~\ref{fig:DCGmonspectr} where the 
monopole cluster length spectrum before and after center vortex removal is 
compared. On the other hand, the non-Abelian string tension vanishes as it 
should~\cite{deforcrand} although only a minimal number of links has actually
been changed. As expected in such a case, the monopole string tension also 
vanishes as it is shown in Fig.~\ref{fig:fullmonabpotentials} where the non-Abelian 
and the monopole potentials before and after vortex removing are depicted.
%---------------------------------------------------------------------
%---------------------------------------------------------------------

%---------------------------------------------------------------------
%---------------------------------------------------------------------

Thus, monopole percolation is not sufficient to generate a confining
potential. The monopole string tension, and consequently the Abelian string
tension, may vanish if the percolating monopole cluster is highly correlated
at small distances. In the following we will present several pieces of numerical
evidence pointing out that the geometrical properties of monopole clusters after
the removal of center vortices in DCG are indeed very different from the 
confining case.

\begin{figure}
\begin{center}
\begin{tabular}{cc}
\includegraphics[width=.45\textwidth,angle=0]{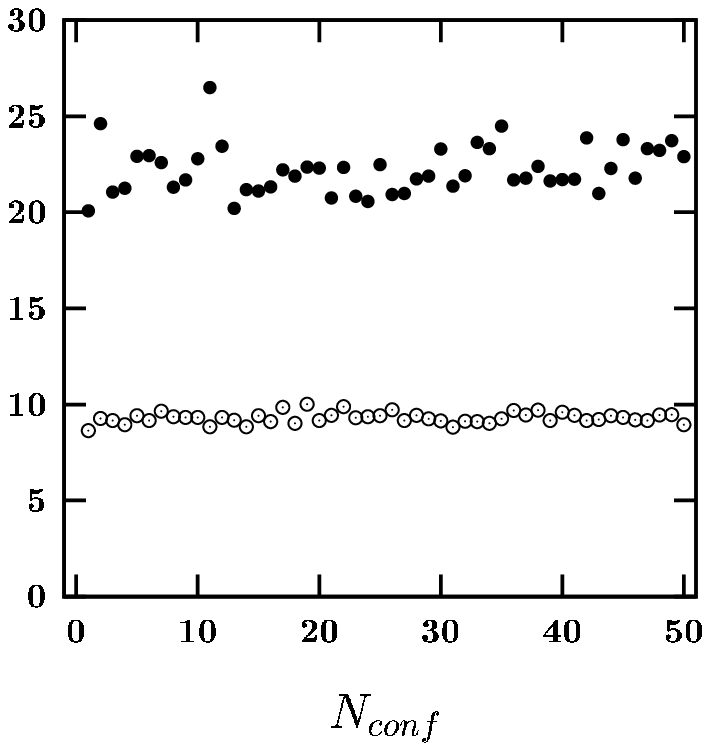}&
\includegraphics[width=.45\textwidth,angle=0]{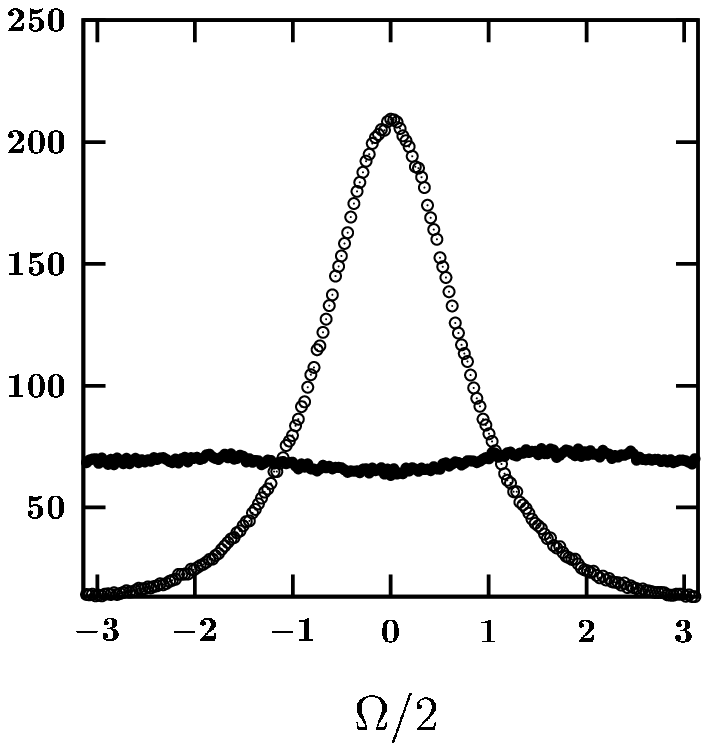}\\
 (a) & (b)
\end{tabular}
\caption{The ratio of the total length of the largest cluster to the
number of crossings in it (a) and the solid angle
$\Omega$ distribution (b) for normal configurations (full circles) and
configurations free of center vortices (empty circles). The error bars in
(b) are less than symbols.}
\label{fig:fig5}
\end{center}
\end{figure}

{\em (i)} The number of selfintersections of the percolating cluster after
removing of center vortices is substantially larger than before, as can
be seen in Fig.~\ref{fig:fig5} a.

%---------------------------------------------------------------------
%---------------------------------------------------------------------

{\em (ii)} A useful quantity to discriminate between confinement and
deconfinement phases at finite temperature is suggested by the following
fact. The monopole part $P^{mon}({\vec x})$ of the Polyakov loop in any 
three-dimensional point ${\vec x}$ is defined in eq. (\ref{eq:monopole-P}).
Using eq. (\ref{eq:monfield}) one can express the monopole part 
$P^{mon}({\vec x})$ as 
\begin{equation}
P^{mon}({\vec x})= e^{i \Omega({\vec x})/2}  \; ,
\end{equation}
where $\Omega({\vec x})$ is the solid angle of all monopole trajectories
(projected onto a single time slice, with proper regard of the direction of
the monopole current) as monitored from the locus ${\vec x}$ of the Polyakov 
loop (thought to be located in the same time slice).

$\Omega({\vec x})$ has been introduced as an observable characterizing the
difference between the confinement and deconfinement phase in Ref.~\cite{Ejiri}.
It has turned out that the distribution of the local values $\Omega({\vec x})$
(the histogram summed over all configurations of an ensemble) is flat in
confinement and has a Gaussian form in the case of deconfinement.
Fig.~\ref{fig:fig5} b shows the distributions of $\Omega({\vec x})$ as found for
normal confining configurations and for the ensemble cleaned from center vortices.
The distribution for the latter is qualitatively very similar to that in the
deconfinement phase.

\begin{figure}
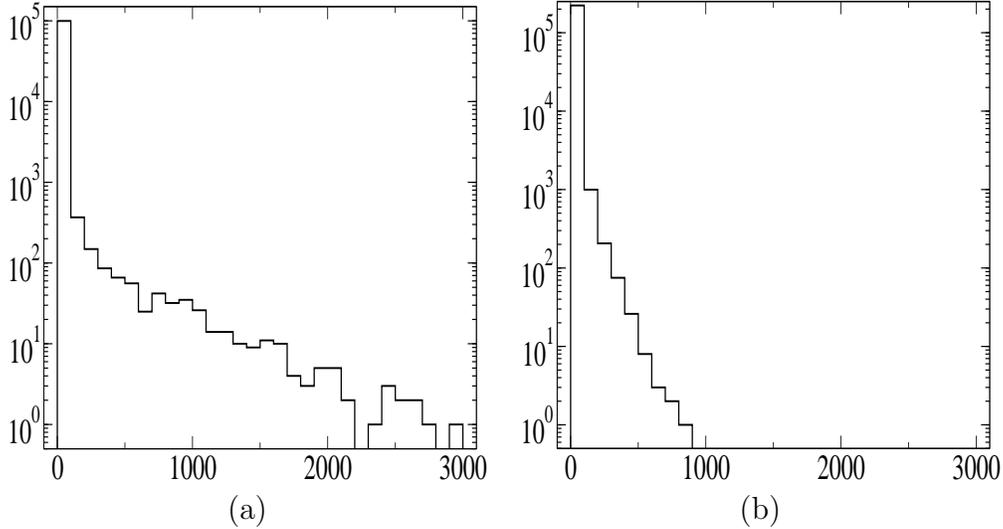

%% [hpbt]
\begin{center}
\begin{tabular}{cc}
\includegraphics[width=.4\textwidth,height=.4\textwidth]{monloopequi.eps}&
\includegraphics[width=.4\textwidth,height=.4\textwidth]{dcgsaloop.eps}\\
(a) & (b)
\end{tabular}
\caption{The length distribution $\tilde{N}(L)$ of irreducible loops emerging 
from the decomposition of monopole clusters, for $\beta=2.35$ on
$24^3 6$ lattices in DCG before (a) and after (b) removing the center
vortices.}
\label{fig:loops}
\end{center}
\end{figure}

{\em (iii)} In order to describe the correlations within the apparently
percolating monopole clusters, it is useful to decompose the cluster by
subtracting from bigger clusters closed, oriented and irreducible 
(non-selfintersecting) loops, 
stepping up in loop size (lengthes of 4, 6, 8 etc.) until nothing remains to
be subtracted. The result of this procedure can be presented as a histogram 
$\tilde{N}(L)$ of irreducible loops (with respect to the loop size) in place 
of the histogram $N(L)$ of connected clusters (with respect to the length of 
monopole currents inside the clusters). 
Fig.~\ref{fig:loops} is the analog to Fig.~\ref{fig:DCGmonspectr}, showing
now the respective {\em loop length distributions}.
It turns out from comparing Figs.~\ref{fig:loops} and \ref{fig:DCGmonspectr}
that the percolating but non-confining monopole clusters can be completely 
decomposed into loops of comparably small size. The irreducible monopole 
loops of the original ensemble (before removing center vortices) ranges to
three times larger loops. Among them are such wrapping around the lattice.

%---------------------------------------------------------------------
%---------------------------------------------------------------------

\begin{figure}
%% [hpbt]
\begin{center}
\begin{tabular}{cc}
\includegraphics[width=.45\textwidth,angle=0]{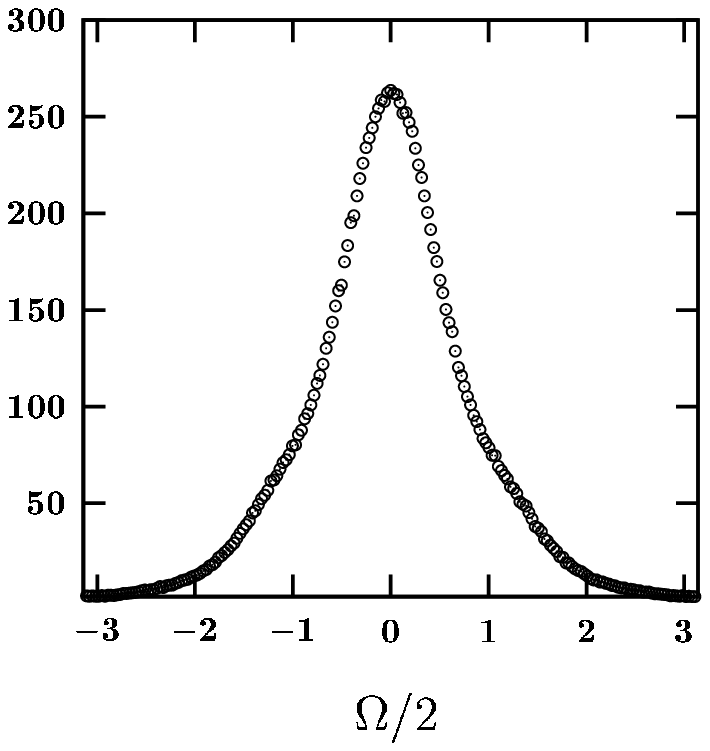}&
\includegraphics[width=.45\textwidth,angle=0]{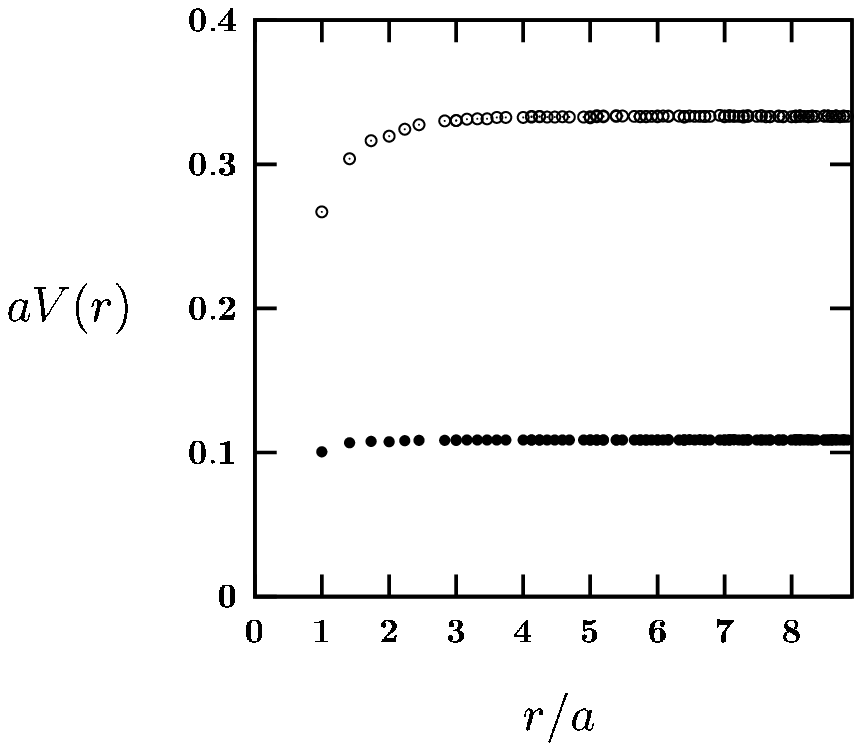}\\
(a) & (b)
\end{tabular}
\caption{The solid angle $\Omega$ distribution (a) for a random monopole 
configurations obtained as described in the text and the static 
monopole-generated 
potential (b) for these configurations (full circles) compared with the 
ensemble free of center vortices (empty circles). Error bars are within 
symbol size.}
\label{fig:fig6}
\end{center}
\end{figure}

This result is corroborated by the following random simulation of non-confining, 
yet percolating monopole clusters. We have tried to model the monopole 
configurations left over after removal of vortices by a random gas of small monopole
loops. Such configurations were created by the following procedure. A set of 
$N_{DPL}$ Dirac plaquettes were put on the lattice with randomly chosen location 
in a randomly chosen plane, with random sign of $m_{x,\mu\nu}$. 
In Fig.~\ref{fig:fig6} the resulting distribution of $\Omega({\vec x})$ and the 
static potential are shown for $N_{DPL}=6,000$. For this number of Dirac plaquettes 
we find as the result of the procedure that a large connected monopole cluster of 
a length of $O(10,000)$ has grown up and the total number of monopole currents 
(including also smaller clusters) is close to that in the P-vortex-removed 
configurations. In Fig.~\ref{fig:fig6} one can see that both the distribution 
of $\Omega({\vec x})$ and the static potential are qualitatively similar to those 
in the ensemble of configurations free of center vortices.

%---------------------------------------------------------------------
%---------------------------------------------------------------------

The facts {\em (i)}, {\em (ii)} and {\em (iii)} show that percolating monopole
clusters, unexpectedly appearing in configurations without P-vortices and without 
confinement, are really highly correlated at small distances and that this can be 
directly related to the lack of confinement. The simulation based on independent 
$1\times 1$ Dirac plaquettes shows that this might be an almost realistic model 
of how these clusters could have been randomly created.

\section{Direct center gauge: removing monopoles}
%------------------------------------------------
\label{sec:DCG_nomono}

\begin{figure}
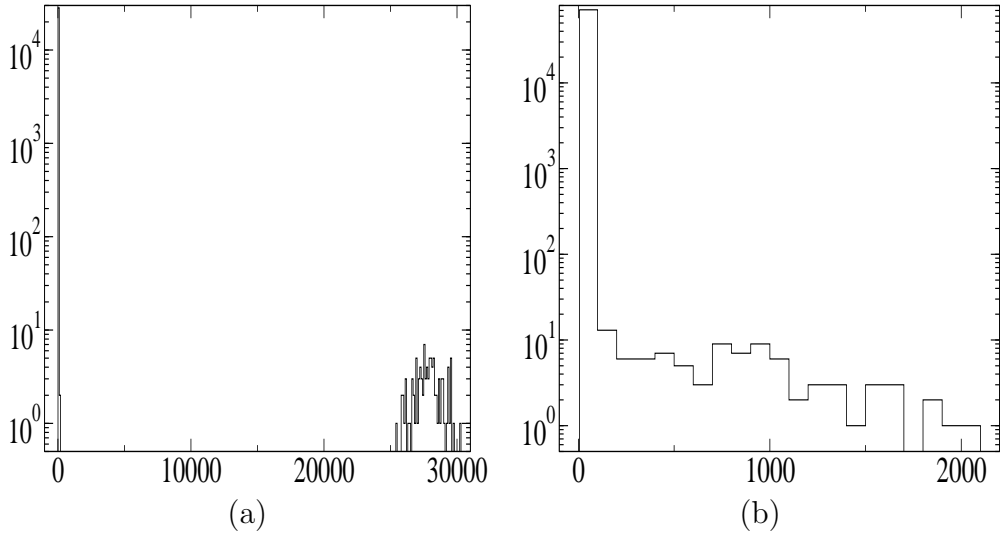

\begin{center}
\begin{tabular}{cc}
\includegraphics[width=.4\textwidth,height=.4\textwidth]{dcgvorclequi.eps}&
\includegraphics[width=.4\textwidth,height=.4\textwidth]{dcgvorclmonrem.eps}\\
(a) & (b)
\end{tabular}
\caption{The same as Fig. \ref{fig:vort_sp} but with vortices defined in DCG,
(a) before and (b) after monopoles have been removed.}
\label{fig:vortex_dcg_nomono}
\end{center}
\end{figure}

Finally, we describe what effect the removal of Abelian monopoles has on the 
P-vortex content in DCG. The basic ensemble of configurations has been put into 
the MAG with the help of the simulated annealing method. From the Abelian projected 
gauge field the singular (monopole) part has been removed before the {\it modified} 
non-Abelian field has been reconstructed using the coset part left out in the 
Abelian projection. This is explained in Eqs. (\ref{eq:monfield}), 
(\ref{eq:mremoved_Abelian}) and (\ref{eq:mremoved_nonAbelian}) in the Appendix. 
Finally we put the two lattice fields into DCG and view the corresponding P-vortex 
content. The P-vortex density $\rho_{\rm vort}=0.0605 \pm 0.0002$ 
({\it i. e.} the total 
vortex area relative to the total number of plaquettes in the lattice) of the 
ensemble with monopoles is reduced to $\rho_{\rm vort}=0.0131 \pm 0.0002$ 
({\it i. e.} roughly one fifth of the original density) in the ensemble where 
monopoles have been removed.
In Fig.~\ref{fig:vortex_dcg_nomono} we compare the area distribution of 
{\it individual} center
vortices after monopole removal (b) with the original distribution obtained
without modifying the gauge fields (a). This Figure is very much similar to 
Fig.~\ref{fig:vort_sp} obtained in ICG. While Fig.~\ref{fig:vortex_dcg_nomono}a
shows a significant part of clusters with very large area consisting of more than
$O(20,000)$ dual plaquettes, after monopole removal such large clusters
do not occur any more. The largest ones have an area less by one order of magnitude
in the number of dual plaquettes and cannot be isolated from the smallest vortices 
occurring in the area distribution. This corresponds to our expectation.

%-------------------------------------------------------------------------------
%---------------------------------------------------------------------

\begin{figure}
\begin{center}
\includegraphics[width=.45\textwidth]{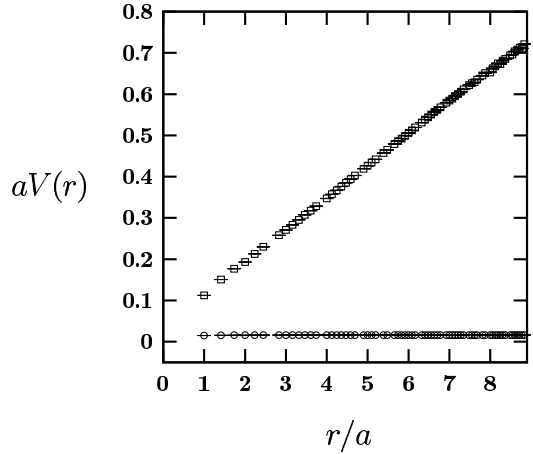}
\caption{The center vortex contribution to the quark-antiquark potential,
rising for the original ensemble put into DCG (denoted by squares),
flat for DCG after Abelian monopoles have been removed (denoted by circles).}
\label{fig:pot_dcg_nomono}
\end{center}
\end{figure}

Fig.~\ref{fig:pot_dcg_nomono} shows the contribution of center vortices to the
quark-antiquark potential in DCG before (upper linearly rising curve) and
after removal of Abelian monopoles (lower flat curve). This picture clearly
shows that the left-over, comparably short-ranged center vortices do not
provide a confining potential.

%-------------------------------------------------------------------------------
%---------------------------------------------------------------------

\section{Conclusions}
%--------------------
\label{sec:Conclusions}

In this paper we have tried to give an answer to the question of the mutual
interdependence of the two types of topological gauge field fluctuations, 
condensation of which is popularly held responsible for quark confinement: 
center vortices (more precisely, P-vortices) and Abelian monopoles. For this 
we have applied techniques of respective removal of one type of these 
fluctuations in order to study the result with respect to the other type. 
Center vortices were identified within the projected $Z(2)$ gauge field obtained
in two ways: {\em (i)} from the indirect center gauge, where the maximally Abelian 
gauge is applied as a first step and the (Abelian) maximal center gauge is fixed 
within the residual Abelian gauge symmetry, before the projection to the center 
subgroup $Z(2)$ is made, and {\em (ii)} through the direct maximally center gauge 
achieved within the non-Abelian gauge freedom.
Abelian monopoles were throughout studied within the maximally Abelian gauge. 
The gauges were iteratively fixed with the help of a simulated annealing algorithm, 
which - when applied to the consecutive step of complimentary gauge fixing - has 
erased the memory with respect to the first step of gauge fixing.

We have found a clear correlation between both types of objects. We confirm that
the removal of each of these types leads to a loss of confinement described in 
terms of the other type, respectively. However, in case of removing P-vortices 
detected within the direct center gauge, we found much more monopole currents  
and larger (spanning) monopole clusters than seen before P-vortex removal. 
A closer inspection showed that these monopole clusters are strongly correlated
at short range such that they are reducible into a large number of considerably 
short monopole loops, the latter not contributing to a linearly rising confinement 
potential. This somewhat surprising observation shows, that spatial percolation 
of infrared monopole clusters alone is not a sufficient condition for {\it real} 
monopole condensation and thus for quark confinement (for the connection of the 
latter see Ref.~\cite{DiGiacomo2,monop_condense}).

We have restricted our investigations to the case of just one temperature
value within the confinement phase. We are convinced that this does not
represent a real loss of generality as long as $T < T_c$.

\section*{Appendix}
%------------------
Here we give the standard definitions for $SU(2)$ lattice gauge theory which
we have used in the studies described in the text. We perform our analyses
both in the Direct~\cite{greensite_3} -- and in the Indirect~\cite{greensite_1} --
Maximal Center Gauges (DCG and ICG). The DCG in $SU(2)$ lattice gauge theory
is defined by the maximization of the functional

\begin{equation}
F_1(U) = \sum_{x,\mu} \left( \tr {}^gU_{x,\mu}\right)^2 \, ,
\label{eq:maxfunc_1}
\end{equation}
with respect to gauge transformations $g \in SU(2)$. $U_{x,\mu}$ is the lattice
gauge field and ${}^gU_{x,\mu}=g^{\dag}(x)U_{x,\mu}g(x+\hat{\mu})$ the gauge
transformed lattice gauge field.
Maximization of (\ref{eq:maxfunc_1}) fixes the gauge up to $Z(2)$ gauge
transformations, and the corresponding, projected $Z(2)$ gauge field is
defined as:
\begin{equation}
Z_{x,\mu} = {\rm sign} \left( \tr {}^gU_{x,\mu} \right) \, .
\label{eq:Zdef}
\end{equation}
After this identification is made, one can make use of the remaining $Z(2)$ gauge
freedom in order to maximize the $Z(2)$ gauge functional
\begin{equation}
F_2(Z) = \sum_{x,\mu} {}^zZ_{x,\mu} \,
\label{eq:maxfunc_2}
\end{equation}
with respect gauge transformations $z(x) \in Z(2)$, 
$Z_{x,\mu} \to {}^zZ_{x,\mu}=z^{*}(x)Z_{x,\mu}z(x+\hat{\mu})$.
This is the $Z(2)$ equivalent of the Landau gauge. In distinction to
Ref.~\cite{deforcrand}, this final step is automatically understood here
before the vortex removal operation (to be defined below) is done.

The ICG goes an indirect way. At first, one fixes the maximally Abelian gauge (MAG)
by maximizing the functional
\begin{equation}
F_3(U) = \sum_{x,\mu} \tr \left( {}^gU_{x,\mu}\sigma_3 ({}^gU_{x,\mu})^{\dag}
\sigma_3\right) \, ,
\label{eq:maxfunc_3}
\end{equation}
with respect to gauge transformations $g \in SU(2)$. The maximization fixes the
gauge up to $g \in U(1)$. Therefore, the following projection to an $U(1)$ gauge
field through the phase of the diagonal elements of the links,
$\theta_{x,\mu} = \arg \left( ({}^gU_{x,\mu})^{11} \right)$, is not unique.
The non-Abelian link field is split according to
$U_{x,\mu}=u_{x,\mu} V_{x,\mu}$ in an Abelian (diagonal) part
$u_{x,\mu} = {\rm diag} \left\{ \exp( i \theta_{x,\mu} ),
                               \exp(- i \theta_{x,\mu} ) \right\}$ and a
coset part $V_{x,\mu} \in SU(2)/U(1)$, the latter representing non-diagonal gluons.
Exploiting the remaining $U(1)$ gauge freedom, which amounts to a shift
$\theta_{x,\mu} \to {}^{\alpha}\theta_{x,\mu}
= - \alpha(x) + \theta_{x,\mu} + \alpha(x+\hat{\mu})$, one can maximize the
Abelian gauge functional
\begin{equation}
F_4(u) = \sum_{x,\mu} \left( \cos( {}^{\alpha}\theta_{x,\mu} ) \right)^2 \, ,
\label{eq:maxfunc_4}
\end{equation}
that serves the same purpose as (\ref{eq:maxfunc_1}).
Finally, the projected $Z(2)$ gauge field is defined as
\begin{equation}
Z_{x,\mu} = {\rm sign} \left( \cos( {}^{\alpha}\theta_{x,\mu} ) \right) \, .
\label{eq:ZdefICG}
\end{equation}
This second step in ICG is completely analogous to the DCG case, in other words
one maximizes $F_1(U)$ in Eq. (\ref{eq:maxfunc_1}), however for all $U_{x,\mu}$ 
restricted to the projected links
$u_{x,\mu} = {\rm diag} \left\{ \exp(  i \theta_{x,\mu} ),
                                \exp(- i \theta_{x,\mu} ) \right\}$,
admitting only $U(1)$ gauge transformations denoted by $\alpha$.
The $Z(2)$ gauge variables are used to form $Z(2)$ plaquettes. Most of them are
equal to $+1$, typically $30,000$ plaquettes are equal to $-1$ after
center projection via DCG (and $40,000$ after center projection via ICG, 
respectively).
The P-vortex surfaces are actually formed by plaquettes {\it dual} to the
negative plaquettes. 
The vortex density is therefore $\rho_{\rm vort} = 0.0605 \pm 0.0002$  
and $\rho_{\rm vort}  = 0.0805 \pm 0.0002$ in DCG and ICG, respectively.

In order to fix the maximally Abelian and the direct maximal center gauge we
have created 10
randomly gauge transformed copies of the original gauge field configuration 
and applied the Simulated Annealing algorithm~\cite{simann} to find the optimal 
non-Abelian gauge transformation $g$. 
We have used for further analyses always {\it that} copy ${}^gU$ 
(or ${}^{\alpha}\theta$) which corresponds to the maximal value of the respective 
gauge fixing functional. In order to perform the second step of the indirect maximal
center gauge via configurations, that have been first fixed to the maximally Abelian
gauge and projected to $U(1)$ fields $u$, we have started again from 10 random 
{\it Abelian gauge transforms} of the latter.

We have used the standard DeGrand--Toussaint definition \cite{DGTmon} of
monopole currents defined by the phase $\theta_{x,\mu}$ of $u_{x,\mu}$.
The typical number of monopole currents per configurations amounts to
$12,000$ dual links, corresponding to a density of 
$\rho_{\rm mon} = 0.0357 \pm 0.0002$.
The part of the Abelian gauge field originating from the monopoles is
\begin{equation}
\theta^{mon}_{x,\mu}  = -2 \pi \sum_{x^\prime} D(x-x^\prime)
\partial_{\nu}^{'} m_{x^{\prime},\nu\mu}\, .
\label{eq:monfield}
\end{equation}
Here $D(x)$ is the inverse lattice Laplacian, and $\partial_{\mu}^{'}$ is the
lattice backward derivative. The Dirac sheet variable, $m_{x,\mu\nu}$, is defined
as the integer multiple of $2\pi$ part of the plaquette angle $\theta_{x, \mu\nu}$,
whereas the reduced plaquette angle $\bar{\theta}_{x,\mu\nu}\in (-\pi,\pi]$ is the
fractional part:
$\theta_{x, \mu\nu} = 2\pi m_{x,\mu\nu} + \bar{\theta}_{x,\mu\nu}$.
The Abelian gauge field with monopoles removed is defined
as~\cite{miyamura_2}:
\begin{equation}
u_{x,\mu}^{{monopole\,\, removed}}=\left( u_{x,\mu}^{mon} \right)^{\dag}
u_{x,\mu}\, ,
\label{eq:mremoved_Abelian}
\end{equation}
where
$u_{x,\mu}^{mon} = {\rm diag} \left\{ \exp(  i \theta^{mon}_{x,\mu} ),
                                           \exp( -i \theta^{mon}_{x,\mu} ) \right\}$.
Upon multiplication with the coset field $V_{x,\mu}$, this holds also for
the non-Abelian links
\begin{equation}
U_{x,\mu}^{{monopole\,\, removed}}=\left( u_{x,\mu}^{mon} \right)^{\dag}
U_{x,\mu}\, .
\label{eq:mremoved_nonAbelian}
\end{equation}
Analogously the gauge fields with the P-vortices removed are defined
as~\cite{deforcrand}:
\begin{equation}
U_{x,\mu}^{{vortex\,\, removed}}=Z_{x,\mu} U_{x,\mu}\, ,
\label{eq:vremoved}
\end{equation}
where $Z_{x,\mu}$ is given by (\ref{eq:Zdef}).

\section*{Acknowledgements}
%--------------------------
The authors want to thank M.N. Chernodub for valuable discussions
and Ph. de Forcrand for useful remarks on a first version of this paper.
The work was partially supported by grants
RFBR-DFG 06-02-04010 and DFG-RFBR 436 RUS 113/739/2.
The ITEP group (A.I.V., B.V.M., M.I.P., V.G.B., A.V.K. and P.Yu.B.)
was partially supported by grants
RFBR 06-02-16309, 05-02-16306, 05-02-17642 and 04-02-16079,
and by the EU Integrated Infrastructure Initiative Hadron Physics (I3HP) 
under contract RII3-CT-2004-506078.
The work of E.-M.I. is supported by DFG through the Forschergruppe 
FOR 465 (Mu932/2).

\end{document}